\documentclass{aa}  

\usepackage{graphicx}
\usepackage{txfonts}
\usepackage{lipsum}
\usepackage{subcaption}       
\usepackage{lscape}            
\usepackage{placeins}          
\usepackage{xcolor}
\usepackage{float}
\usepackage{academicons}
\usepackage{xcolor}
\usepackage{orcidlink}
\usepackage{hyperref}
\hypersetup{colorlinks=True,citecolor=blue,allcolors=blue}
                                
\usepackage{hyperref}
\hypersetup{colorlinks=true,linkcolor=blue,citecolor=blue,urlcolor=blue}

\newcommand{\rev}[1]{#1}
\newcommand{\revv}[1]{#1}

\definecolor{darkmagenta}{rgb}{0.55, 0.0, 0.55}
\begin{document}

   \title{Physically motivated AGN emissivity profiles and their effects on quasar microlensing signatures}
   \subtitle{I. Multi-epoch accretion disc size inference}

   \author{Scott Hagen\orcidlink{0000-0002-5075-7920}\inst{1, 2, 3}
        \and Carina Fian\orcidlink{0000-0002-2306-9372}\inst{3, 4}
        }

   \institute{
   IFPU - Institute for Fundamental Physics of the Universe, Via Beirut 2, 34151 Trieste, Italy\\
    \email{shagen@sissa.it}
    \and SISSA - International School for Advanced Studies, Via Bonomea 265, 34136 Trieste, Italy 
     \and Center for Astrophysics and Cosmology, University of Nova Gorica, Vipavska 11c, 5270 Ajdov\v{s}\v{c}ina, Slovenia \\
    \email{carina.fian@ung.si}
    \and INAF - Osservatorio Astronomico di Trieste, Via G. B. Tiepolo 11, I-34143 Trieste, Italy
   }

  \abstract
  {
  Quasar microlensing is uniquely sensitive to the size-scale of the accretion flow, offering one of the few direct probes of the accretion structure on micro-arcsecond scales. However, microlensing-based measurements in the optical and ultraviolet (UV) often find sizes systematically larger than expected from standard Shakura-Sunyaev disc theory, commonly referred to as the `disc-size problem' similar to that seen in continuum reverberation campaigns. But this assumes that all the emission comes from a single compact disc, neglecting the diffuse emission from the broad-line region (BLR) which originates on much larger spatial scales. In this paper we directly quantify the effect of large-scale diffuse emission on the observed microlensing signatures. We adapt the physically motivated {\sc agnsed} model to construct energetically self-consistent emissivity profiles in any given bandpass, for both standard discs and the modified warm Comptonised discs. Since this also predicts the full spectral energy distribution (SED), we combine these SEDs with {\sc cloudy} to give a diffuse BLR component under the assumption of photo-ionised equilibrium. We convolve these models with representative microlensing magnification maps, and generate mock microlensing light curves to directly assess the inferred source size under different physical conditions. 
  While the detailed shape of the disc emissivity profile has only a higher-order effect on the microlensing profile, the inclusion of the BLR makes a significant impact since this naturally smooths out the caustic network over larger scales. This introduces a significant bias when interpreted purely as a compact disc. However, the strength of this bias depends predominantly on the fractional contribution of the diffuse emission to the SED in the bandpass being considered, as this sets the effective half-light radius, giving an important wavelength dependence. We conclude that part of the excess in microlensing-inferred accretion disc sizes could arise from interpreting a composite (disc+BLR) picture as a single compact disc.
  }


  \titlerunning{AGN emissivity profiles and their effects on microlensing signatures}
   \authorrunning{S. Hagen \& C. Fian}

   \maketitle
    \nolinenumbers

\section{Introduction}
Active galactic nuclei (AGN) are powered by accretion onto supermassive black holes (SMBHs), often described by the standard \citet{Shakura73} disc model. While the power released in this accretion flow gives rise to some of the most luminous objects in the Universe \citep[e.g.][]{LyndenBell69}, the relevant physical size scales are far too small to image directly. Instead, we rely on indirect methods, such as reverberation mapping \citep[e.g.][]{Blandford82,Peterson93,Edelson19} and gravitational microlensing \citep[e.g.][]{Chang1979}, to map the structure of the accretion flow. In this paper, we focus on the latter.

Gravitational microlensing is uniquely sensitive to the physical size of the background source. 
In strongly lensed quasars, microlensing arises when stars and other compact objects in the lensing galaxy differentially magnify the unresolved micro-images of a background quasar (\citealt{Chang1979,Chang1984,Wambsganss2006}, \rev{see also the recent review by \citealt{Vernardos2024}}). Because these micro-images are not spatially resolved, microlensing manifests as flux variability that probes source structure on sub-parsec scales.

As the caustic network produced by foreground stars sweeps across the accretion flow, the resulting magnification depends on both the size and surface-brightness profile of the emitting region, making microlensing a powerful tool for probing the inner structure of AGN. However, accretion-disc sizes inferred from microlensing statistics \citep[e.g.][]{Morgan2010,Blackburne2011,Jimenez2012,Fian2016,Fian2018,Fian2021} are often systematically larger than predicted by standard \citet{Shakura73} theory. In most microlensing analyses, the source surface-brightness distribution is simplified to facilitate the modelling, commonly using circular Gaussian profiles or standard thin-disc prescriptions \citep[e.g.][]{Mortonson2005,Jimenez2014}. These choices capture the leading-order dependence of the microlensing signal on source size \citep{Mortonson2005}, but they neglect more detailed emissivity profiles and the presence of extended continuum-emitting structures such as winds or the broad-line region (BLR). Such components may produce distinctive microlensing signatures and systematic biases in the inferred disc sizes \citep{Fian2023_diffuse}.

A similar disc-size tension is also seen in intensive black-hole reverberation mapping (IBRM) campaigns, which often infer disc sizes a factor of $\sim 2$--$4$ larger than expected from the standard disc model \citep[e.g.][]{Edelson19,Fian2022_RM,Fian2023_RM,Kara2023,Lewin2024,Mandal2025}. These campaigns also reveal a pronounced lag excess around $\sim 3000$\,\AA\ \citep[e.g.][]{Cackett18,Edelson19,Hernandez20}, commonly referred to as the U-band excess. Both the long lags and the U-band excess can be explained by additional reprocessing from the BLR or from a wind launched near the inner BLR \citep[e.g.][]{Korista01,Korista19,Lawther18,Dehghanian19b,Chelouche19,Kara21,Netzer22,Hagen24a}.

The BLR is expected to subtend a large solid angle as seen from the central source, with a covering fraction of order $\sim 0.3$ \citep[e.g.][]{Czerny11,Baskin18}, and therefore absorbs a significant fraction of the ionizing extreme-ultraviolet (EUV) spectral energy distribution (SED). Recombination of ionized gas in the BLR produces diffuse free-bound continuum emission \citep{RybickiLightman86,Korista01,Korista19}, which contributes to the optical/UV portion of the SED. Since the BLR is located on physically much larger scales than the accretion disc, this diffuse component naturally produces longer time lags when measured relative to a relatively clean disc-continuum band, i.e. one in which the diffuse-continuum contribution is minimal.

An analogous effect is expected for microlensing. Diffuse free-bound continuum emission modifies the effective optical/UV surface-brightness profile by adding flux from physically larger BLR scales. Since microlensing magnification depends on the spatial distribution of the emitted light, this extended component dilutes the microlensing signal from the compact accretion disc and changes the predicted magnification distributions, as shown by \citet{Fian2023_diffuse}. Building on that work, we extend the analysis to physically motivated AGN emissivity profiles by comparing standard thermal-disc and warm-Comptonisation models, computing the diffuse BLR continuum from the ionizing SED, and testing how these composite source models bias disc-size recovery in multi-epoch microlensing observations.

We base our intrinsic source models on the {\sc agnsed} framework of \citet{Kubota18}, which divides the accretion flow into a standard outer disc, an intermediate warm Comptonised disc, and an inner hot flow. This model captures two important departures from a simple standard disc: the ubiquitous X-ray power-law tail, generally attributed to hot Comptonisation \citep{Haardt93, Haardt94}, and the optical/UV--soft-X-ray connection associated with warm Comptonisation \citep[e.g.][]{Done12,Kubota18, Petrucci13, Petrucci18}. For microlensing, these components are relevant because they affect both the compact optical/UV emissivity profile and the ionizing EUV continuum that powers the diffuse BLR emission. We use the {\sc agnsed} framework to construct two-dimensional emissivity maps for two fiducial intrinsic continua: a standard \citet{Shakura73} disc and a warm-Comptonised disc. We then pass the corresponding SEDs through the {\sc cloudy} photoionisation code \citep{Ferland17,Gunasekera25} to compute the diffuse continuum emission from BLR-scale gas, providing a physically motivated framework for assessing how intrinsic disc emissivity profiles and extended BLR continuum emission affect microlensing magnification distributions and inferred disc sizes. We use these models to generate mock microlensing observations, which we then forward model to assess the resulting size-recovery, allowing for a direct estimation of the biases induced by neglecting extended structures or alternative emissivity profiles. We find that, in general, microlensing size recovery is primarily sensitive to the effective half-light radius of the observed surface-brightness distribution, while the detailed compact-disc emissivity profile plays a secondary role. The addition of diffuse BLR continuum emission can therefore bias compact-only size estimates by increasing the effective continuum size and diluting the microlensing variability. The magnitude of this bias is wavelength dependent and is set by both the diffuse-continuum flux fraction and the compact-disc emissivity profile used as the reference. Thus, by analogy with IBRM campaigns, part of the microlensing disc-size tension \rev{could possibly arise} from interpreting composite disc+BLR emission as emission from a single compact accretion disc.

This paper is organized as follows. In Section~\ref{sec:emissivitymodels}, we describe the physical SED and disc emissivity models. In Section~\ref{BLRmodel}, we present the BLR model, including the diffuse continuum contribution. Section~\ref{MLpred} focuses on the microlensing predictions, while in Section~\ref{SizeImpact} we use mock microlensing observations to examine how different emissivity profiles affect inferred source sizes. Finally, in Section~\ref{conclusions}, we summarize our main conclusions.

\section{Physical SED and disc emissivity models}\label{sec:emissivitymodels}

\subsection{Model implementation}
We begin by defining the underlying AGN SED model and describing how it is translated into an emissivity map for a given bandpass. We base the SED on the radially stratified {\sc agnsed} model of \citet{Kubota18}. While we refer the reader to \citet{Kubota18} for a detailed description of {\sc agnsed}, we provide a brief overview of the key concepts and of our implementation for completeness. Throughout this work, we use the standard notation for radius, where $R$ denotes the radius in physical units and $r$ denotes the dimensionless radius in gravitational radii. These quantities are related by $R=r\,R_{\rm G}$, where $R_{\rm G}=GM/c^2$, $G$ is the gravitational constant, and $c$ is the speed of light. Similarly, for the mass-accretion rate, we use $\dot{M}$ to denote the physical mass-accretion rate and $\dot{m}$ to denote the dimensionless Eddington-scaled mass-accretion rate. These are related by $\dot{M}=\dot{m}\dot{M}_{\rm Edd}$, where $\dot{M}_{\rm Edd}=L_{\rm Edd}/[\eta(a)c^2]$ is the Eddington mass-accretion rate and $\eta(a)$ is the spin-dependent radiative efficiency.

We assume that the entire flow follows a \citet{Novikov73} emissivity profile, such that the temperature profile scales as $T_{\rm NT}^4(R) \propto R^{-3} f(R)$, where $f(R)$ describes the local disc structure and is derived in the Kerr metric by \citet{Page74}. When computing both the SED and the emissivity maps, we divide the flow into a series of geometrically spaced radial annuli, each with width $\Delta R$ and temperature $T_{\rm NT}(R)$. Throughout this paper, the radial resolution is fixed to 500 bins per radial decade in $R_{\rm G}$.

Following \citet{Kubota18}, we radially stratify the accretion flow into three distinct regions: an outer standard \citet{Shakura73} disc for $r_{\rm out} \geq r > r_{\rm w}$, a warm Comptonised disc for $r_{\rm w} \geq r > r_{\rm h}$, and an inner hot X-ray plasma, also referred to as the hot Comptonisation region, for $r_{\rm h} \geq r > r_{\rm isco}$.

For the standard-disc region, we assume that the disc is fully thermalised, such that each annulus emits as a black body with spectrum $B_{\nu}(T_{\rm NT}(R))$. The bolometric luminosity of an annulus is then given by $2 \times 2\pi R \Delta R \sigma T_{\rm NT}^{4}(R)$, where $\sigma$ is the Stefan--Boltzmann constant. The leading factor of $2$ accounts for the disc having two sides.

For radii $r_{\rm w} \geq r > r_{\rm h}$, we assume that the disc does not fully thermalise, forming instead a warm Comptonisation region, or warm corona. This picture is motivated by the work of \citet{Petrucci13, Petrucci18, Petrucci20}, in which the disc vertical structure differs from the standard \citet{Shakura73} prescription \citep[see, e.g., the simulations of][]{Rozanska15, Jiang20, Kawanaka24}. In this region, a significant fraction of the accretion power is dissipated in the photosphere rather than the disc mid-plane. The resulting dissipative layer can be described as a warm, $kT_{\rm e} \sim 0.3$\,keV, optically thick, $\tau \gg 1$, plasma forming a slab-like geometry above a passive underlying disc. The power dissipated in this plasma irradiates and is reprocessed by the underlying disc, heating it and producing seed photons, which subsequently Compton scatter through the warm plasma. Following \citet{Kubota18}, we tie the seed-photon temperature to that expected from the \citet{Novikov73} solution, such that $T_{\rm seed}(R)=T_{\rm NT}(R)$. Each annulus is therefore assigned the two-sided luminosity $4\pi R\Delta R \sigma T_{\rm NT}^{4}(R)$. Internally, we model the spectral shape of each annulus using {\sc nthcomp} \citep{Zdziarski96, Zycki99}. We further assume that the entire warm Comptonising plasma has a single electron temperature, $kT_{\rm e,w}$, and a covering fraction of unity as seen from the underlying disc. The photon index, $\Gamma_{\rm w}$, is left as a free parameter, since it depends on the uncertain heating-cooling balance of the warm plasma. This also allows for cases in which part of the accretion power is dissipated in the underlying disc, as suggested at higher $\dot{m}$ \citep[e.g.][]{Petrucci18, Kubota18, Chen_SJ25}. We note, however, that an optically thick corona with a covering fraction of unity is expected to have $\Gamma_{\rm w} \gtrsim 2.5$ \citep{Petrucci18}, with $\Gamma_{\rm w}\sim2.5$ corresponding to the case of a truly passive underlying disc.

Below $r_{\rm h}$, we assume that the disc truncates and is replaced by an optically thin, $\tau \sim 1$, geometrically thick, $H/R \sim 1$, hot, $kT_e\sim100$\,keV, plasma \citep{Narayan95, Liu99, Rozanska00b}. We refer to this as the hot Comptonisation region, or hot corona. As in \citet{Kubota18}, the plasma is assumed to be powered by the accretion flow between $r_{\rm h}$ and $r_{\rm isco}$, while still following the \citet{Novikov73} emissivity profile. This gives the dissipated luminosity, $L_{\rm diss}$. The seed photons that are Compton scattered in this region originate from the outer disc and/or the warm Comptonisation region. These seed photons also contribute to the total power output of the hot Comptonised plasma through $L_{\rm seed}$, such that the total hot-corona luminosity is $L_{\rm h}=L_{\rm diss}+L_{\rm seed}$. We refer the reader to \citet{Kubota18} for details of how these quantities are calculated. As for the warm Comptonisation region, we compute the spectral shape using {\sc nthcomp}. The seed photons are expected to be dominated by emission from the inner edge of the disc, since this is both where the disc is brightest and where the solid angle subtended by the corona, as seen from the disc, is largest. We therefore set the seed-photon temperature to $T_{\rm NT}(R_{\rm h})\exp(y_{\rm w})$, where $y_{\rm w}$ is the Compton $y$-parameter of the warm Comptonisation region. In the absence of a warm Comptonisation region, the seed-photon temperature is simply $T_{\rm NT}(R_{\rm h})$. We further assume a constant electron temperature, $kT_{\rm e,h}$. The photon index, $\Gamma_{\rm h}$, is left as a free parameter, although we note that it should in principle be determined by the heating--cooling balance of the plasma when photon-number conservation is taken into account, which is not included here.

The above prescription gives the intrinsic emission in {\sc agnsed}, and is so far identical to the implementation of \citet{Kubota18}. However, the hot-corona emission is assumed to be isotropic, so a fraction of the emitted X-rays should be incident on the disc and heat it \citep[e.g.][]{Zycki99, Cackett07}. This irradiation modifies the effective disc temperature in both the outer standard-disc region and the warm Comptonisation region according to
$\sigma T_{\rm eff}^{4}(R) = \sigma T_{\rm NT}^{4}(R) + F_{\rm rep}(R)$, where $F_{\rm rep}(R)$ is the reprocessed X-ray flux incident on the disc annulus at radius $R$. At this point, we depart slightly from \citet{Kubota18}, who included irradiation of only one side of the disc. Here, instead, we allow the disc to be heated on both sides, following the implementation of \citet{Hagen23a}.

The total SED is then obtained by computing the emitted spectrum from each annulus and summing over all annuli. The key point, however, is that these models also retain the relevant spatial information. This property has been used extensively to model AGN reverberation and variability \citep{Gardner17, Mahmoud20, Mahmoud23, Hagen23a, Hagen24a}. Here, we build on this framework to generate 2D emissivity maps for microlensing studies, by further subdividing each annulus azimuthally with a resolution of $\Delta \phi = 0.05$. Crucially, because the full SED is calculated, emissivity maps can be extracted for any chosen bandpass. We note that we currently neglect the effects of general-relativistic light bending by the central black hole. While these effects are certainly present, they are concentrated to the innermost regions. In the context of {\sc agnsed}, they predominantly affect the X-ray emission \citep[as shown by][]{Hagen23b}, which is not the focus of this current paper. In the optical/UV, general-relativistic effects are typically small, although this depends to some extent on the disc truncation radius and black-hole spin \citep{Dovciak22, Hagen23b}. 
For this study the dominant effect would be a reduction of the observed frame emissivity from small radii, due to light bending out of the observer's line of sight. Given that microlensing is predominantly sensitive to the physical size of the source at larger radii \citep{Mortonson2005} we expect these corrections to be completely negligible.

\begin{figure}
    \centering
    \includegraphics[width=0.88\columnwidth]{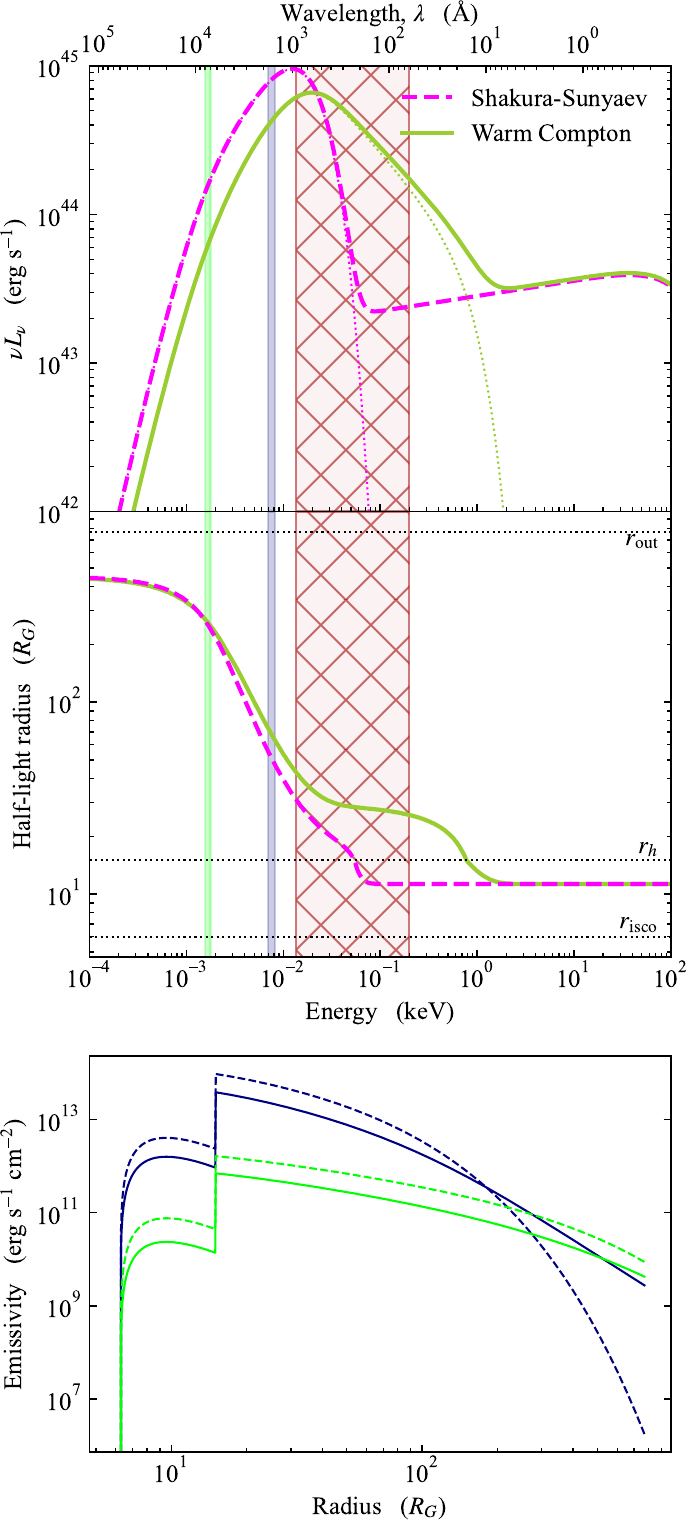}
\caption{
    \textbf{\textit{Top:}} Model SEDs for the fiducial warm-Comptonised disc model (green solid line) and standard \citet{Shakura73} disc model (magenta dashed line). In both cases, the accretion flow truncates into an inner X-ray-emitting plasma below $r_h=15$, producing the high-energy X-ray tail. The dotted curves show the individual disc and warm-Comptonised disc components. The hatched region marks the unobservable energy range, and the shaded regions indicate the bandpasses used to extract the radial emissivity profiles shown in the bottom panel. \textbf{\textit{Middle:}} Half-light radius as a function of spectral energy for the two fiducial models. The horizontal dotted lines mark characteristic radii, including $r_{\rm isco}$, $r_h$, and $r_{\rm out}$, as labelled. \textbf{\textit{Bottom:}} Radial emissivity profiles extracted in the highlighted bandpasses. Solid lines correspond to the warm-Comptonised disc model, while dashed lines correspond to the standard \citet{Shakura73} disc model. The colours match the shaded bandpasses in the top panel. The sharp drop at $15\,R_{\rm G}$ marks the transition radius where the flow truncates into the inner X-ray-emitting plasma.
}
    \label{fig:sed_em_rhl}
\end{figure}

\subsection{Example SEDs and emissivity}
Here, we present two example SEDs computed using the implementation described above, and with input parameters typical of those inferred for nearby AGN \citep[e.g.][]{Jin12a, Jin12c, Mitchell23}. Both models adopt the same black-hole parameters, but differ in their assumed disc structure. SED\,1 corresponds to a pure \citet{Shakura73} disc extending from the self-gravity radius, $r_{\rm sg}$ \citep[calculated following][]{Laor89}, down to $r_{\rm h}=15$, where the disc truncates into the inner hot corona. SED\,2 instead corresponds to a disc fully covered by a warm Comptonising corona, also extending from $r_{\rm sg}$ down to $r_{\rm h}=15$. In both cases, we assume a black hole mass of $M=10^8\,M_{\odot}$ and an Eddington-scaled accretion rate of $\dot{m}=0.1$. We fix the black-hole spin to $a=0$, for which the impact of the neglected general-relativistic ray-tracing effects is expected to be small \citep{Hagen23b}. We also fix the hot-corona parameters to $kT_{\rm e,h}=100$\,keV and $\Gamma_{\rm h}=1.9$ for both SEDs. For SED\,2, we set the warm-corona parameters to $kT_{\rm e,w}=0.2$\,keV and $\Gamma_{\rm w}=2.7$. These two SEDs form our fiducial example models, used throughout this paper and are shown in Fig.\,\ref{fig:sed_em_rhl}. Throughout this work, all quantities are given in the AGN rest-frame unless stated otherwise.

In the warm Comptonisation model, the electrons in the plasma have, on average, higher energies than the seed photons, so the emitted spectrum is preferentially Compton up-scattered \citep{Petrucci13, Petrucci18, Kubota18}. In the SED, this shifts the emission to higher energies, with repeated scatterings producing a continuum that bridges the EUV region \citep{Done12, Kubota18}. More important for microlensing, however, is the effect on the emissivity profile in a given bandpass, shown in the bottom panel of Fig.\,\ref{fig:sed_em_rhl}. Here, we consider two bandpasses: a high-energy band centred at $\lambda_{\rm mid}=1650$\,\AA\ with width $\Delta\lambda=250$\,\AA, and a low-energy band centred at $\lambda_{\rm mid}=7500$\,\AA\ with width $\Delta\lambda=1000$\,\AA. These bands are chosen to be well separated in wavelength, to highlight the energy dependence of the emissivity profile, and they correspond broadly to the wavelength ranges commonly probed in optical/UV studies. As shown in the next section, these two bands also receive very different contributions from the BLR diffuse-recombination continuum. 

The bottom panel of Fig.\,\ref{fig:sed_em_rhl} shows the resulting radial emissivity profiles in our two filters for both models, with solid and dashed lines corresponding to the warm Compton and \citet{Shakura73} models, respectively. The drop at small radii, seen for all models and filters, marks the truncation of the disc into the inner hot X-ray plasma. This has a negligible impact on the microlensing signatures considered here. At larger radii, the profiles begin to differ in a way that depends strongly on wavelength. For the low-energy filter, the emissivity profile is relatively shallow in both models. This is because this bandpass lies close to, or on, the Rayleigh--Jeans tail of the local emission, where at fixed frequency $B_{\nu}\propto T \propto R^{-3/4}$. This radial dependence is largely preserved in the warm Comptonised model, since inverse Comptonisation retains the shape of the low-energy tail of the incident spectrum. The main difference is therefore a change in normalisation: because Comptonisation shifts power to higher energies, the low-energy band samples further down the Rayleigh--Jeans tail, resulting in a lower overall emissivity. For the high-energy filter, the emissivity profile instead reflects the transition between different spectral regimes of the local emission. This is especially apparent in the Shakura-Sunyaev model, where the profile rises steeply at large radii and then flattens towards smaller radii. This behaviour can be understood as a transition from the Wien tail of the local black-body spectrum at large radii to the vicinity of the local black-body peak at smaller radii. The warm Comptonised model shows a more pronounced change in shape because Compton up-scattering redistributes the local emission into a broader, approximately power-law-like spectrum, with the Wien-like cut-off shifted to energies above the warm-corona electron temperature, i.e. into the soft X-ray regime. As a result, an optical/UV filter does not sample the Wien tail of the local emission at any radius, leading to a flatter observed emissivity profile.

The main implication is that these two models share the same underlying temperature profile, namely that of the \citet{Novikov73} solution, but can nevertheless produce very different emissivity profiles over specific wavelength ranges. This difference arises entirely from the change in emission mechanism, from thermal black-body emission to thermal Comptonisation. The effect is strongest at energies for which $kT(R_{\rm out}) \ll E$, where the outer disc contributes through the Wien tail in the thermal model. At lower energies, both models sample the Rayleigh--Jeans tail over much of the disc, and therefore have similar radial emissivity shapes. This behaviour is naturally reflected in the predicted half-light radius, shown in the middle panel of Fig.\,\ref{fig:sed_em_rhl}. Along the main disc continuum, the half-light radius of the warm Compton model decreases more slowly with increasing energy than in the standard-disc case. It then flattens in the extreme-UV, once the band lies beyond the peak of the local Comptonised spectrum, although this regime is difficult to observe directly for most AGN.

The disc emission alone, however, does not capture the full spatial structure of the observed continuum. AGN also contain material with large-scale heights, typically associated with the BLR and/or winds. This gas intercepts part of the central extreme-UV emission and reprocesses it into diffuse free-bound emission \citep[e.g.][]{Korista01, Korista19}. Since this component contributes to the observed SED, it also modifies the spatial emissivity profile and therefore the expected microlensing signatures. We account for this component explicitly in the next section.

\begin{figure}
    \centering
    \includegraphics[width=0.93\columnwidth]{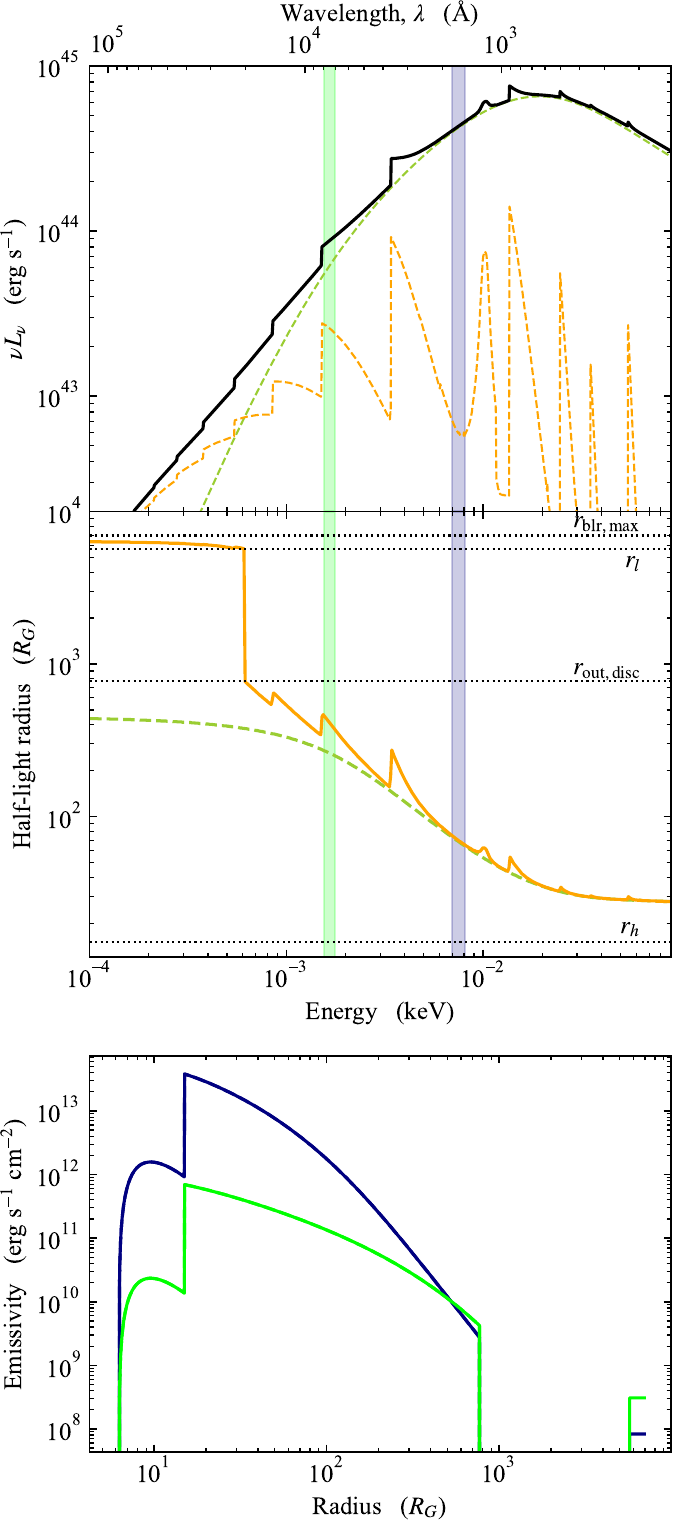}
   \caption{
    \textbf{\textit{Top:}} Optical/UV portion of the model SED including the diffuse BLR continuum contribution. The intrinsic warm-Comptonised disc emission is shown as a dashed green line, the diffuse BLR component as a dashed orange line, and the total observed SED as a solid black line. The shaded regions indicate the bandpasses used to extract the radial emissivity profiles shown in the bottom panel and later used to compute the microlensing magnification distributions. \textbf{\textit{Middle:}} Half-light radius as a function of wavelength for the intrinsic warm-Comptonised disc alone (dashed green line) and for the combined disc+BLR model (solid orange line). The dotted horizontal lines mark characteristic model radii, as labelled. \textbf{\textit{Bottom:}} Radial emissivity profiles for the two highlighted bandpasses, including the extended BLR contribution at large radii. The colours correspond to the shaded wavelength regions in the top panel.
}
    \label{fig:blr_sed}
\end{figure}
\section{Implementing the BLR emission and geometry}\label{BLRmodel}
\label{sec:blr_geom}

Having introduced the intrinsic SED and emissivity profiles, we now include the diffuse continuum emission from BLR-scale gas and assess its impact on the predicted microlensing distributions. We use {\sc cloudy} v25.00 \citep{Ferland17,Gunasekera25} to calculate the diffuse recombination continuum produced by constant-density gas illuminated by the intrinsic SEDs described in Section~\ref{sec:emissivitymodels}. Our approach is similar to that of \citet{Hagen24a}, who used {\sc cloudy} simulations to quantify the impact of diffuse continuum emission on AGN variability and continuum reverberation lags.

In this work, we simplify the problem by considering only time-averaged SEDs. As in \citet{Hagen24a}, we assume a simple bi-conical BLR geometry, described in Appendix~\ref{app:blr_goem}. The geometry is defined by the launch radius, $r_l$, the launch angle measured from the disc plane, $\alpha_l$, and the total covering fraction as seen from the central source, $f_c := \Omega/4\pi$. The ionization balance in {\sc cloudy} is set by the absolute luminosity of the incident SED and the distance between the central source and the gas. For simplicity, we represent the BLR using a single {\sc cloudy} calculation evaluated at the geometric mean radius between $r_l$ and $r_{\rm BLR,max}$, where $r_{\rm BLR,max}$ is determined by $\alpha_l$ and $f_c$. This assumes a constant gas density and a single ionization state across the BLR. Although this is a simplified description, more complex models including gradients in density, radius, and ionization state \citep[e.g.][]{Korista19,Lawther18} are beyond the scope of the present work. Our aim here is not to construct a detailed BLR model, but to quantify the first-order impact of diffuse recombination continuum emission on microlensing magnification distributions and inferred source sizes.

We demonstrate the BLR contribution using the fiducial warm-Comptonised SED introduced in the previous section. We assume a covering fraction of $f_c=0.3$, motivated by the predictions of \citet{Baskin18}. The launch radius is chosen such that the geometric mean BLR radius, $r_{\rm BLR,mid} := 10^{0.5\log_{10}(r_l r_{\rm BLR,max})}$, lies on the BLR radius--luminosity relation of \citet{Bentz13}, using the 5100\,\AA\ luminosity of our fiducial warm-Comptonised SED. For a launch angle of $\alpha_l=60^\circ$, this gives $r_l=5700\,R_{\rm G}$. For the {\sc cloudy} calculation, we adopt a column density of $N_{\rm H}=10^{23}$\,cm$^{-2}$ and a constant hydrogen number density of $n_{\rm H}=10^{11.5}$\,cm$^{-3}$. These values yield mostly neutral material at the BLR mid-radius, with $\log_{10}\xi=-0.28$ for the adopted input SED.

Because the quasars considered here are unobscured type-I AGN, they are expected to be viewed at relatively low or moderate inclinations within the standard orientation-based unification picture \citep{Antonucci1993,Urry1995}. 
We therefore assume that the observer views the system along the axis of the bi-cone rather than through the BLR material. Under this geometry, we use the reflected {\sc cloudy} spectrum, which includes both diffuse emission from the illuminated face of the cloud and backscattered incident continuum. We assume that this emission is radiated isotropically from the BLR surface into $2\pi$\,sr and that the BLR surface emissivity is uniform. The projected two-dimensional BLR surface brightness is then obtained by dividing the observed BLR luminosity by the projected surface area of the BLR annulus, as described in Appendix~\ref{app:blr_goem}. We further assume that only the BLR emission from the side of the disc facing the observer is visible, so that the observed BLR luminosity is half of the total luminosity calculated by {\sc cloudy}.

The resulting optical/UV spectrum, half-light radius spectrum, and radial emissivity profiles are shown in Fig.~\ref{fig:blr_sed}. For clarity, we show only the line-subtracted diffuse continuum. The effect of the BLR on the half-light radius and normalized emissivity profile is closely related to its fractional contribution to the SED at a given wavelength. Including line emission would further increase the BLR contribution in bandpasses contaminated by broad emission lines, thereby increasing both the effective half-light radius and the weight of the extended component in the microlensing convolution. 

In terms of the absolute surface emissivity, shown in the bottom panel of Fig.~\ref{fig:blr_sed}, the diffuse BLR continuum appears modest even in the bandpass where its contribution to the SED is significant. This apparent discrepancy is a consequence of the large emitting area of the BLR. Although the BLR surface emissivity per unit area is much lower than that of the compact disc, the BLR covers a large annular region between $r_l$ and $r_{\rm BLR,max}$. Its integrated emission can therefore make a substantial contribution to the total SED, the half-light radius, and the resulting microlensing magnification distributions.

\section{Microlensing predictions}\label{MLpred}
In this section, we assess the impact of each source model on the predicted microlensing magnification distributions. We begin by generating two representative source-plane magnification maps, $I(x,y)$, as described in Section~\ref{sec:magmap_generation}. Each map gives the microlensing magnification that would be measured for a point source located at each position in the source plane, for a fixed macrolens configuration and realization of the stellar microlens population. The macrolens describes the smooth, large-scale lensing by the foreground galaxy, which sets the image positions, parities, and macro-magnifications, while the microlenses produce the small-scale caustic structure within each image. The map values are expressed as relative flux magnifications, $F/F_0$, where $F_0$ denotes the flux expected in the absence of microlensing, or equivalently after normalization by the corresponding macro-model magnification. The distribution of values in the unconvolved map defines the point-source microlensing magnification distribution, which we use as the reference case before accounting for the finite extent of the source. Since quasars are spatially extended, the magnification distribution relevant for an observed source is obtained by convolving the source-plane magnification map with the normalized two-dimensional surface-brightness profile of the emitting region,
\begin{equation}
    I_{\rm obs}(x,y) = (I \circledast \hat{\epsilon})(x,y),
\end{equation}
where $\hat{\epsilon}(x,y)$ is the model AGN emissivity profile normalized to unity total flux. This convolution smooths the small-scale caustic structure in the magnification map, with the degree of smoothing depending on the size and shape of the source emissivity profile. We examine the resulting model-dependent magnification distributions in Section~\ref{sec:mod_dist}. 

\subsection{Microlensing magnification maps}
\label{sec:magmap_generation}

The statistical properties of the source-plane magnification maps, $I(x, y)$, are determined by the local convergence ($\kappa$) and shear ($\gamma$) at the position of each lensed image. The convergence quantifies the projected surface mass density and can be decomposed as $\kappa = \kappa_{\mathrm{c}} + \kappa_{\star}$, where $\kappa_{\mathrm{c}}$ accounts for smoothly distributed matter (e.g., dark matter) and $\kappa_{\star}$ represents the contribution from compact objects such as stars.  The shear $\gamma$ governs the anisotropy of the microlensing caustic network, stretching the magnification pattern along one axis and compressing it along the orthogonal direction. High shear produces elongated, filamentary caustics, whereas low shear leads to more isotropic structures. Throughout this work, we assume a stellar mass fraction of approximately 20\% of the total surface mass density near the Einstein radius (\citealt{Jimenez2015}).

Microlensing magnification maps in the source plane were generated using the Fast Multipole Method–Inverse Polygon Mapping (FMM–IPM; see \citealt{Jimenez2022}) technique\footnote{\url{https://gloton.ugr.es/microlensing/}}, which efficiently combines fast deflection-angle calculations of \citet{Greengard1987} with high-resolution polygon mapping of \citet{Mediavilla2006,Mediavilla2011}. Each map spans a region of $20\,r_E$ on a side, where \rev{we adopt a typical} $r_E \approx 3\times10^{16}$ cm \rev{\citep[e.g][]{Mosquera2011}}. This is sampled on a $4000 \times 4000$ pixel grid, corresponding to a pixel scale of $0.058\,\rm{light\text{-}days}\rev{\, \approx 10.2\,R_G}$. \rev{This is smaller than the characteristic UV-emitting scales explored in our simulations, which extend from $\sim15\,R_G$ outward.} \rev{The stellar component of the lens was modelled as a random distribution of microlenses with individual masses of $M_\star=0.3\,M_\odot$, a representative value commonly adopted in quasar microlensing studies \citep[e.g.][]{Jimenez2012}.} While the detailed caustic morphology depends on the adopted stellar mass function, the resulting magnification probability distributions are known to be largely insensitive to this choice (\citealt{Lewis1995,Wyithe2001,Chan2021}). For alternative microlens masses, source sizes can be rescaled according to the relation $r_E \propto M_\star^{1/2}$, since the projected Einstein radius sets the natural physical length scale of the magnification maps \citep[e.g.][]{Wambsganss2006,Kochanek2004,Morgan2010}. The magnification value at each pixel is normalized to the mean macro-model magnification of the corresponding image.

\begin{figure*}
    \centering
    \includegraphics[width=0.99\textwidth]{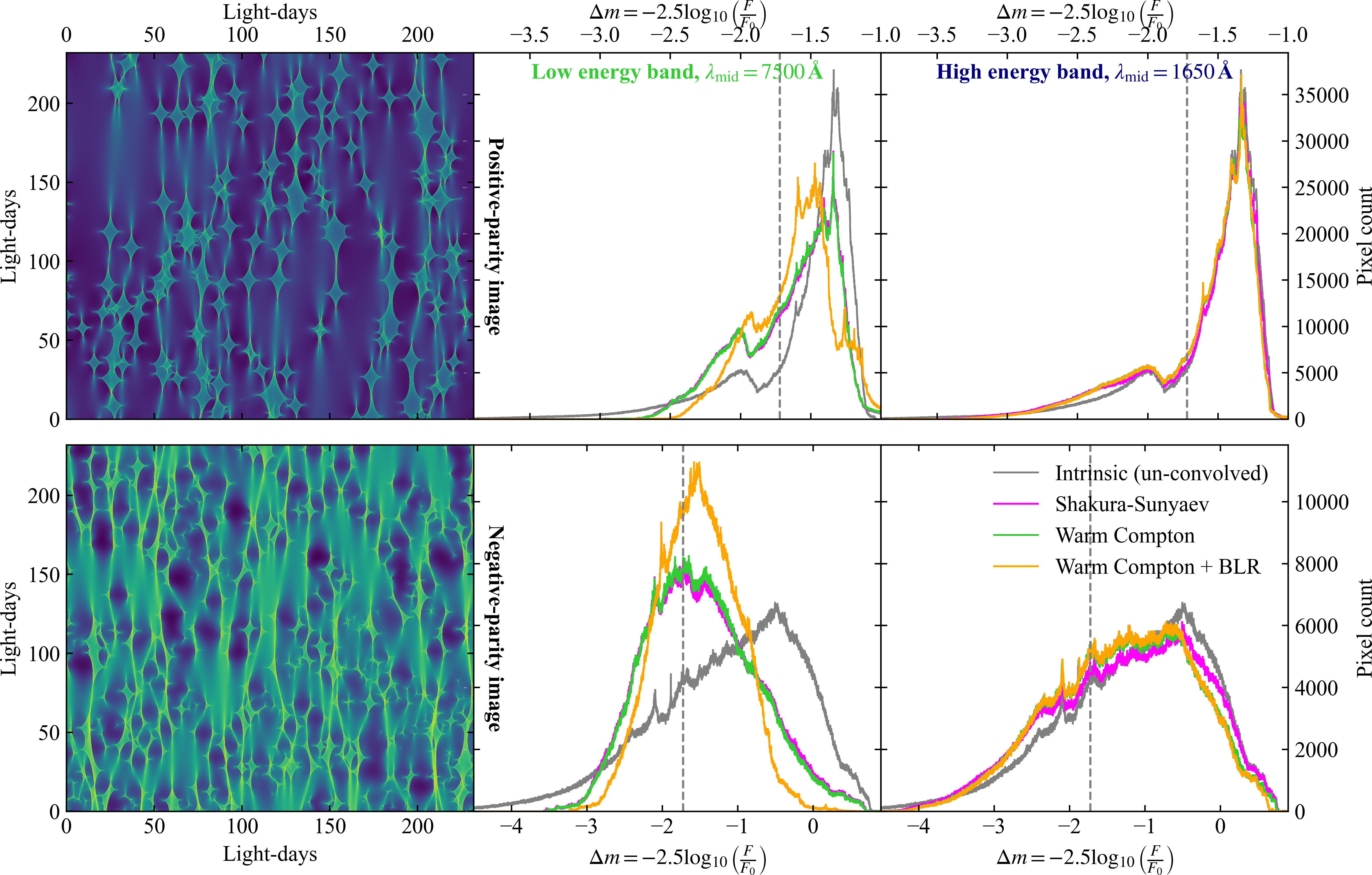}
    \caption{
    Model convolution results for the positive-parity image (top row) and negative-parity image (bottom row).
    \textit{\textbf{Left column:}} Unconvolved source-plane magnification maps used to generate the example histograms on the right. These maps describe the expected magnification at each source-plane position and are convolved with the model 2D emissivity profiles to obtain the extended-source magnification maps.
    \textbf{\textit{Middle and right columns:}} Magnification histograms calculated from the source-plane magnification maps after convolution with the source models. We show the unconvolved point-source case (grey), the standard \citet{Shakura73} disc (magenta), the warm-Comptonised disc model (green), and the warm-Comptonised disc plus diffuse BLR continuum model (orange). The models correspond to those described in Sections~\ref{sec:emissivitymodels} and \ref{BLRmodel}. The histograms are shown for two fiducial filters: one centred at $\lambda_{\rm mid}=7500$\,\AA\ (middle column) and one centred at $\lambda_{\rm mid}=1650$\,\AA\ (right column). These filters were chosen because they have very different fractional BLR contributions and correspond to the bandpasses highlighted in Figs.~\ref{fig:sed_em_rhl} and \ref{fig:blr_sed}. \revv{The dashed vertical grey lines show the mean magnification (calculated in flux units) of the unconvolved maps, which agree with the expected theoretical value within $2.5\%$ and $2.3\%$ for the positive and negative parity images respectively.} \rev{We note that the plotted $\Delta m$ values retain the constant macro-magnification offset of the adopted lens configuration and are therefore not centred at $\Delta m=0$.}
    }
    \label{fig:model_hist}
\end{figure*}

To capture the main microlensing behaviours, we adopt a fiducial quasar setup based on two representative magnification maps, both shown in Fig.\,\ref{fig:model_hist}, corresponding to the two generic image parities encountered in strongly lensed quasars. One map represents a positive-parity (minimum-type; $\kappa=\gamma= 0.4$) image, while the other corresponds to a negative-parity (saddle-point; $\kappa=\gamma= 0.6$) image; these parities arise from the local properties of the lensing potential and are known to produce systematically different caustic morphologies and microlensing statistics (\citealt{Schechter2002}). Positive-parity images typically exhibit smoother magnification patterns, whereas negative-parity images are characterized by denser and more asymmetric caustic networks, leading to enhanced microlensing variability. By analysing both configurations, we span the dominant regimes of quasar microlensing and assess how source surface-brightness structure influences microlensing observables in a way that is largely independent of the details of individual lens systems.

\subsection{Microlensing magnification distributions}
\label{sec:mod_dist}

We convolve the source-plane magnification maps with the model emissivity profiles described in Sections~\ref{sec:emissivitymodels} and \ref{BLRmodel}. Although the intrinsic source models are computed on a two-dimensional grid, the fiducial configurations considered here are axisymmetric. The corresponding two-dimensional surface-brightness profile can therefore be understood as the radial emissivity profiles shown in Figs.~\ref{fig:sed_em_rhl} and \ref{fig:blr_sed} rotated around the symmetry axis.

Before the convolution, each emissivity map is normalized to unity, such that:
\begin{equation}
    \hat{\epsilon}(x,y) = \frac{\epsilon(x,y)}{\sum_{x,y} \epsilon(x, y)} \approx
    \frac{\epsilon(x,y)\,\mathrm{d}A_p}
    {\int_S \epsilon(x,y)\,\mathrm{d}A}
\end{equation}

where $dA_p$ is the area of one pixel in the magnification map and the denominator is the total flux emitted over the source plane. With this normalization, the convolved map $I_{\rm obs}(x,y) = (I \circledast \hat{\epsilon})(x,y)$
represents the microlensing magnification that would be observed for an extended source whose centre is located at position $(x,y)$ on the map.

After convolution, we construct the model microlensing distributions by histogramming the convolved map values in terms of $\Delta m = -2.5\log_{10}(F/F_0)$, where $F/F_0$ is the relative microlensing magnification. We choose units of $\Delta m$ to remain consistent with observational studies. These histograms describe the probability of observing a given microlensing amplitude for a source with the assumed surface-brightness profile and lensing configuration. In a multi-epoch monitoring campaign, the relative motion of the source, lens, observer, and microlenses samples different regions of the magnification map. The observed distribution of microlensing amplitudes therefore provides a statistical realization of the underlying model distribution, with the degree of convergence depending on the length and sampling of the campaign.

The resulting distributions for each source model and for both fiducial filters are shown in Fig.~\ref{fig:model_hist}. The effect of the diffuse BLR continuum is most pronounced in the low-energy bandpass, centred at $\lambda_{\rm mid}=7500$\,\AA, which overlaps the Paschen continuum. In this band, the BLR contributes a substantial fraction of the total flux, $f_{\rm BLR}\sim30\%$. Since this emission arises over the much larger BLR annulus, between $R_l$ and $R_{\rm BLR,max}$, it acts as an extended smoothing component when convolved with the magnification map. As a result, the sharp caustic structure is partially averaged out, reducing the variance of the convolved map and producing a narrower, more strongly peaked magnification distribution.

This behaviour can be understood as a consequence of the relative spatial scales involved. The compact disc component is sensitive to the fine structure of the caustic network and is therefore more likely to populate the high- and low-magnification tails of the distribution. By contrast, the extended BLR component averages over many small-scale caustic features, suppressing these tails. The magnitude of this effect depends not only on the BLR flux fraction, but also on the ratio between the characteristic BLR size and the dominant spatial scales of variation in the magnification map.

The high-energy bandpass, centred at $\lambda_{\rm mid}=1650$\,\AA, shows a much weaker BLR effect. In this case, the diffuse BLR contribution is small, $f_{\rm BLR}\sim1.4\%$, and the warm-Comptonised disc and warm-Comptonised disc+BLR models therefore produce nearly identical magnification distributions. Moreover, the emission in this band is concentrated at smaller radii, with half-light radii of order sub light-day for both the warm-Comptonised and standard-disc models. These compact source profiles smooth the caustic network much less efficiently than the extended BLR component, and their magnification distributions therefore remain closer to the unconvolved point-source case.

\section{Size inference from simulated light curves}\label{SizeImpact}
In this section, we generate mock microlensing light curves and use them to recover source sizes from variability statistics, allowing us to quantify the \rev{the bias that arises when the recovered half-light radius of a composite disc+BLR source is interpreted as the size of the compact accretion disc alone.}

\subsection{Mock light curve generation}

\begin{figure*}
    \centering
    \includegraphics[width=\textwidth]{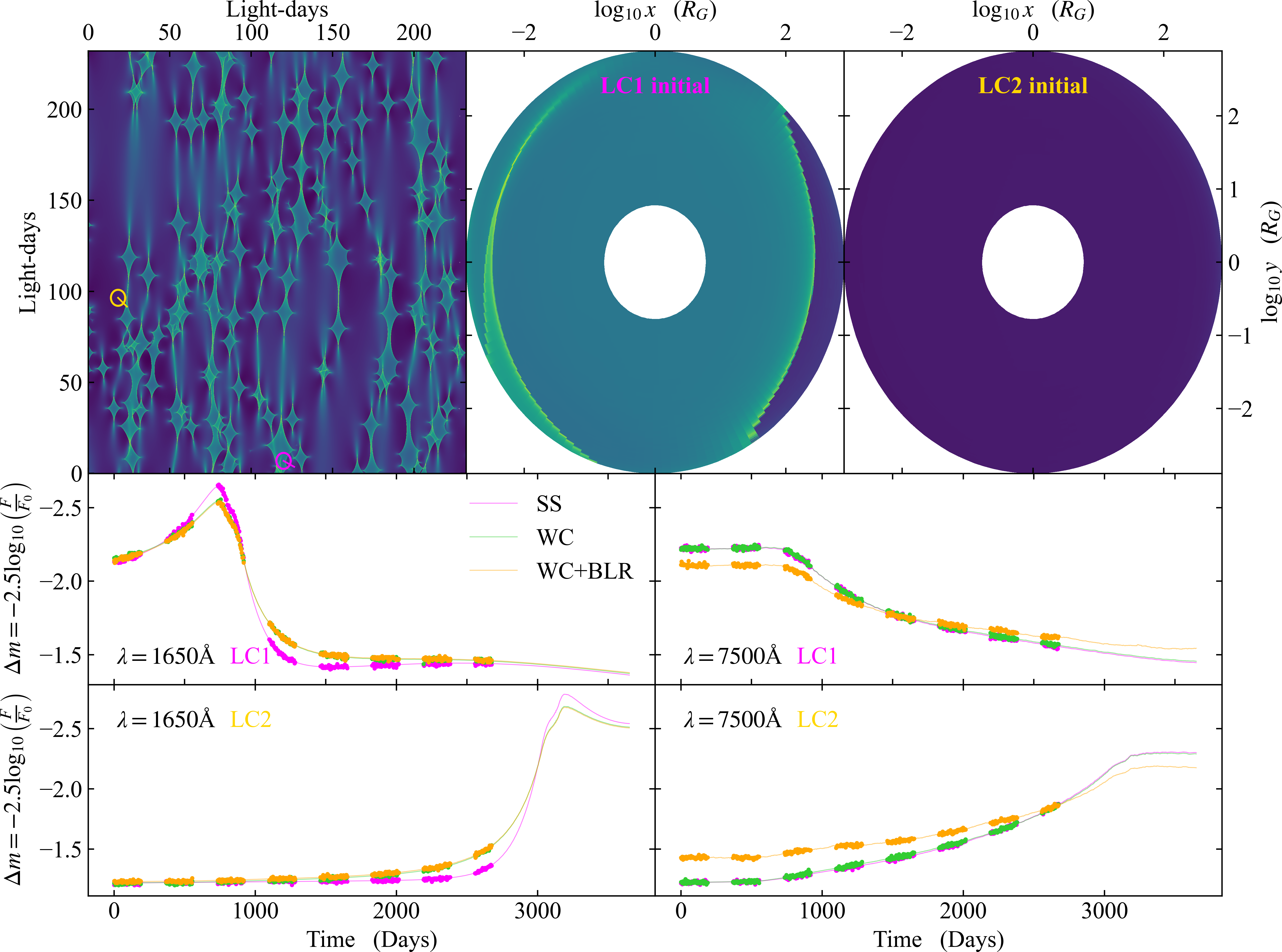}
    \caption{
    Example simulated light curves. \emph{\textbf{Top row:}} The left panel shows the input magnification map for the positive-parity image used to generate the light curves. The coloured circles, shown in magenta and yellow, mark the initial disc positions, while the corresponding straight lines indicate the direction of motion. The middle and right panels show the magnification pattern across the AGN model grid used to compute the magnified emissivity profile for light curve 1 and light curve 2, respectively. \emph{\textbf{Middle and bottom rows:}} The corresponding simulated light curves are shown for light curve 1 in the middle row and light curve 2 in the bottom row. The left and right columns show the light curves extracted in the high- and low-energy bands, respectively, at $\lambda=1650$\,\AA\ and $\lambda=7500$\,\AA. This illustrates the wavelength dependence of the emissivity profile and of the resulting microlensing magnification. The two examples were intentionally selected to correspond to tracks passing through very different regions of the magnification map, demonstrating that the absolute microlensing signal is highly sensitive to the source position on the map and therefore motivating a statistical inference approach. The solid lines show the output model tracks, while the data points show the light curves after re-sampling, adding noise, and adding seasonal gaps.}
    \label{fig:simLCs}
\end{figure*}

Microlensing variability arises from the relative motion between the source, the lensing galaxy, and the observer, including the motions of individual stars in the lensing galaxy. A microlensing light curve can therefore be generated by drawing a trajectory on the source-plane magnification map, moving the source along this trajectory, and computing the magnification across the AGN model grid at each position. This yields a position-dependent source magnification profile, which is multiplied by the intrinsic emissivity profile to obtain the magnified 2D emissivity profile. The observed luminosity is then calculated from the surface integral of the magnified emissivity profile. In practice, the AGN model grid does not coincide exactly with the pixel grid of the source-plane magnification map. We therefore treat each position along the trajectory as the centre of the AGN grid, corresponding to the location of the black hole. The magnification map is then sampled onto the AGN grid by assigning to each AGN grid point the value of the source-plane magnification pixel containing the centre of that point. Example interpolated 2D magnification profiles are shown in the top middle and right panels of Fig.\,\ref{fig:simLCs}.

For simplicity, we consider linear tracks whose lengths are set by the effective transverse velocity of the source on the source plane, $v_{\rm eff}$, and by the total light curve duration. We adopt a fiducial value of $v_{\rm eff} = 600$\,km\,s$^{-1}$, based on the peculiar velocity dispersion inferred by \citet{Mediavilla2016}. To define an observational set-up (i.e., noise, duration, and cadence) representative of current monitoring data, we characterize the publicly available {\sc cosmograil} light curves (\citealt{Eulaers2013,Kumar2013,Courbin2018,Bonvin2018,Bonvin2019,Millon2020b,Millon2020a}). For systems with observations from multiple telescopes, light curves of the same lensed quasar image were merged before computing the global sample properties. 
The resulting {\sc cosmograil} sample consists of 34 lensed quasars and 89 observed image light curves.
\rev{These correspond to individual lensed images and should not be interpreted as 89 independent microlensing light curves; for a system with $n$ images, only $n-1$ independent differential microlensing light curves can be constructed.} The median light curve contains 325 epochs and spans a temporal baseline of approximately 7 yr 4 months. The median spacing between consecutive observations is 4.5 days, reflecting the typical cadence during the observing seasons rather than a uniform sampling over the full baseline, which also includes seasonal visibility gaps and other interruptions in the monitoring. 
The median photometric uncertainty is 0.008 mag. We adopt these values as representative observational constraints when generating mock microlensing light curves. The temporal sampling and photometric noise are therefore fixed to {\sc cosmograil}-like values, while the underlying microlensing signal is generated by drawing trajectories across the source-plane magnification map. 

To sample a broad range of possible caustic-crossing signatures, we draw 100 paths with randomized starting positions and directions. \rev{The choice $N_{\rm LC}=100$ was adopted as a compromise between sampling the stochastic dependence on source trajectories and computational cost.} For each path, we generate microlensing light curves for all three example SED models considered in this work: Shakura-Sunyaev, warm Compton, and warm Compton+BLR. 
This is done for three wavelength bands: the two fiducial bands shown previously, centred on $\lambda=1650$\,\AA\ and $\lambda=7500$\,\AA, as well as an additional one centred on $\lambda=3250$\,\AA\ with a width of $\Delta \lambda=500$\,\AA\, thereby sampling the Balmer continuum.
For each simulated light curve, we assign independent photometric uncertainties by drawing error bars from a Gaussian distribution centred on the median {\sc cosmograil} uncertainty of 0.008\,mag, with a width of 0.0008\,mag. These uncertainties are treated as $1\sigma$ errors. Noise is then added to each light curve point by drawing from a Gaussian distribution centred on zero, with a standard deviation given by the corresponding error bar. Finally, the light curves are re-sampled to a {\sc cosmograil} like cadence and include seasonal observing gaps.
These simulated light curves form the basis of the size-recovery analysis presented below. Example light curves for two independent paths and for each SED model are shown in Fig.\,\ref{fig:simLCs}, illustrating how the local region of the magnification map sampled by the source can strongly affect the observed microlensing variability.

\subsection{Size recovery from variability statistics} \label{sec:size_recovery_statistics}
Microlensing source sizes can, in principle, be inferred through detailed forward modelling of the observed light curves, in which trial trajectories are fitted directly to the data for a given lens model, microlens population, source profile, and effective transverse velocity \citep[e.g.,][]{Kochanek2004}. This approach is powerful, but computationally expensive because it requires exploring a high-dimensional parameter space of source sizes, trajectories, velocities, and microlens configurations. As an alternative, several microlensing studies have used statistical approaches in which observed microlensing amplitudes or variability measures are compared with distributions predicted from magnification maps convolved with sources of different sizes \citep[e.g.,][]{Guerras2013,Fian2016,Fian2018,Fian2021,Rojas2020,Fores2024}. Here we adopt this second approach, since our goal is not to model a specific observed system, but to quantify how different emissivity prescriptions bias the size inferred from multi-epoch microlensing variability.

For each assumed emissivity model $M$, we define a logarithmically spaced grid of trial half-light radii, $r_{1/2}$. For each trial size and bandpass, we first compute the native model half-light radius in that band, $r_{1/2,\,\rm native}$, and define the scale factor
\begin{equation}
a \equiv \frac{r_{1/2}}{r_{1/2,\,\rm native}} .
\end{equation}
The trial source profile is then obtained by rescaling the physical coordinate grid of the fixed dimensionless emissivity pattern by this factor. This changes the physical source size in the source plane while leaving the normalized surface-brightness profile unchanged. 
The procedure should therefore be understood as a controlled size rescaling rather than as a new accretion-flow calculation: we do not recompute the SED, temperature structure, or wavelength-dependent emissivity profile for a different black-hole mass. Its purpose is to test how the microlensing variability changes when the physical half-light radius of a given source morphology is varied.

For each trial size, we convolve the rescaled source profile with the source-plane magnification map and sample the same set of $N_{\rm LC}=100$ mock trajectories described above. From each simulated light curve, we measure the standard deviation of the microlensing variability,
\begin{equation}
\sigma_{\Delta m} =
\left[
\frac{1}{N-1}
\sum_{j=1}^{N}
\left(\Delta m_j-\overline{\Delta m}\right)^2
\right]^{1/2},
\end{equation}
where $N$ is the number of epochs and $\overline{\Delta m}$ is the mean microlensing magnitude along the track. This statistic provides a measure of the microlensing variability amplitude. Compact sources sample the small-scale caustic network more efficiently and therefore produce larger values of $\sigma_{\Delta m}$, whereas extended sources smooth the magnification pattern and produce weaker variability. 

For a given emissivity model $M$ and trial half-light radius $r_{1/2}$, the ensemble of simulated tracks defines the distribution of variability amplitudes expected for that source size,
\begin{equation}
p(\sigma_{\Delta m} \mid r_{1/2}, M),
\end{equation}
which we estimate using a one-dimensional kernel-density estimate. This distribution provides the calibration between source size and microlensing variability for the adopted emissivity model. For each mock observed light curve, with measured variability amplitude $\sigma_{{\rm obs},i}$, we evaluate this distribution at the observed value. The likelihood of a trial source size is therefore
\begin{equation}
\mathcal{L}_i(r_{1/2}\mid M)
=
p(\sigma_{{\rm obs},i}\mid r_{1/2},M).
\end{equation}

Rather than averaging individual best-fitting sizes, we combine the likelihoods from all mock light curves to construct a joint posterior,
\begin{equation}
p_{\rm joint}(r_{1/2}\mid M) \propto
p(r_{1/2})
\prod_{i=1}^{N_{\rm LC}}
\mathcal{L}_i(r_{1/2}\mid M).
\end{equation}
We adopt a prior that is flat in $\log r_{1/2}$, that is, $p(r_{1/2}) \propto 1/r_{1/2}$ over the simulated size range. The resulting joint posterior averages over the stochastic dependence on the particular trajectories through the magnification map and provides the statistical size estimate expected from a {\sc cosmograil}-like ensemble of microlensing light curves. A schematic sketch illustrating the main steps of this size-recovery procedure is provided in Appendix~\ref{size_recovery}.

This framework allows us to recover the most likely half-light radius from mock multi-epoch microlensing variability and to quantify systematic biases introduced by the assumed emissivity model. In the following analysis, we generate mock observations from the composite warm-Comptonisation+BLR source model and then recover the source size using simplified compact-only emissivity prescriptions. This allows us to test how interpreting composite disc+BLR emission as a single compact source biases the inferred microlensing size.

\subsection{\rev{Effect of diffuse BLR emission on inferred disc sizes}}

\begin{figure}
    \centering
    \includegraphics[width=\columnwidth]{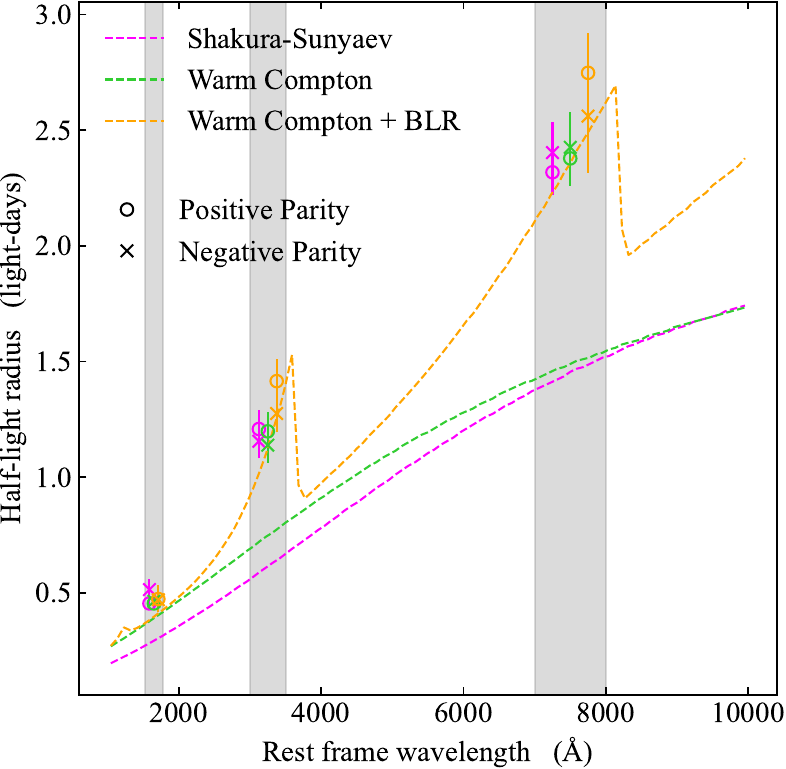}
    \caption{
    Recovered half-light radii from simulated microlensing light curves generated using the composite warm-Comptonisation+BLR source model. The size recovery is performed under three different modelling assumptions: a Shakura--Sunyaev disc (magenta), a warm-Comptonisation disc (green), and the full warm-Comptonisation+BLR model (orange), for both positive- and negative-parity magnification maps. The shaded regions indicate the filters used to generate the mock microlensing light curves. In all cases, the recovered sizes tend toward the effective half-light radius of the input warm-Comptonisation+BLR model, shown by the dashed orange curve. For completeness, we also show the half-light radii expected for the compact Shakura-Sunyaev and warm-Comptonisation models, indicated by the dashed magenta and green curves, respectively, highlighting the apparent size overestimation introduced by the diffuse BLR contribution. \rev{The recovered half-light radii are obtained from the joint posterior constructed by combining the likelihoods of $N_{\rm LC}=100$ mock light curves. The relatively small uncertainties therefore reflect the ensemble constraint; the size inferred from an individual mock light curve is substantially less precise.}
    }
    \label{fig:inferredSize}
\end{figure}

Fig.\,\ref{fig:inferredSize} shows the recovered sizes, expressed as half-light radii, compared with the expected half-light radii of the input source models. These recovered sizes were obtained from mock microlensing light curves generated using the composite warm-Comptonisation+BLR model, while performing the size recovery under each of our three fiducial emissivity assumptions: a Shakura-Sunyaev disc, a warm-Comptonisation disc, and the full warm-Comptonisation+BLR model. This allows us to directly assess the bias introduced when composite disc+BLR emission is interpreted using compact-only source prescriptions. 

It is clear from Fig.\,\ref{fig:inferredSize} that, regardless of the emissivity profile assumed in the size recovery, the inferred sizes tend toward the effective half-light radius of the input composite warm-Comptonisation+BLR model. In Appendix\,\ref{recovered_sizes} we give the full set of recovered sizes for each input model used to simulate the light curves. These same-model tests show that the recovery procedure returns sizes consistent with the input half-light radii to the accuracy required for the comparisons made here. This confirms that microlensing is primarily sensitive to the characteristic source size, and only secondarily to the detailed emissivity profile, as already suggested by \citet{Mortonson2005}. The more important implication is that part of the microlensing disc-size tension may arise from reprocessed continuum emission on BLR scales. This is physically plausible, since the BLR is known to contribute not only to the broad emission lines but also to the continuum \citep{Korista01,Korista19}, with clear evidence for Balmer-continuum contributions in some nearby AGN \citep[e.g.][]{Mahmoud20,Mehdipour15}. Our results show that this additional BLR-scale continuum contribution can significantly alter the effective half-light radius inferred from microlensing.

\rev{This interpretation is also supported by recent observational microlensing studies. \citet{Fian2023_diffuse} showed that including diffuse BLR continuum emission can help reproduce the excess continuum sizes and size--wavelength relations inferred from published microlensing measurements of lensed quasars. More recently, \citet{Hutsemekers2025}  simultaneously constrained the continuum and BLR sizes in eight lensed quasars and found that, on average, the inner radius of the BLR begins near the outer extent of the UV--optical continuum-emitting region. They also found a correlation between BLR and continuum size, supporting the presence of an extended continuum component associated with the BLR. These results provide independent observational motivation for the composite compact-disc+extended-continuum picture explored here.}

The apparent size overestimation can be quantified by comparing the half-light radius of the composite disc+BLR source with that of the corresponding compact-only disc component at the same wavelength. Naively, one might expect this effect to behave like a simple flux-weighted average, similar to the way diffuse BLR emission affects continuum reverberation lags \citep{Hagen24a}. However, for microlensing half-light radii the situation is more subtle. The monochromatic half-light radius is defined by
\begin{equation}
    \label{eqn:r_half}
    2\pi \int\limits_{R_{\rm in}}^{R_{1/2}} \epsilon_{\nu}(R)R\,\mathrm{d}R = \frac{1}{2}L_{\nu} = \pi \int\limits_{R_{\rm in}}^{R_{\rm out}} \epsilon_{\nu}(R)R\,\mathrm{d}R
\end{equation}

Here, $\epsilon_{\nu}(R)$ denotes the total monochromatic emissivity profile of the observed source, including both the compact disc and any extended diffuse BLR continuum component.

In the limit where the BLR contributes more than half of the observed flux in a given bandpass, the composite half-light radius will move onto BLR scales, such that $R_{1/2}\gtrsim R_{\rm l}$. This can be more than an order of magnitude larger than the compact-disc half-light radius. For the BLR SED shown in Fig.\,\ref{fig:blr_sed}, however, this regime occurs only at $\lambda \gtrsim 10^4$\,\AA, where strong microlensing signatures are not generally expected because of the increasing importance of host-galaxy and torus emission. The more relevant case for quasar microlensing is therefore the optical/UV range, where the diffuse BLR continuum fraction is typically below 50\% for the continuum-only models considered here, although it may be larger in bandpasses affected by broad emission lines.

In this optical/UV regime, the composite half-light radius remains inside the outer disc radius for the models considered here. The BLR contribution then enters mainly through the total luminosity on the right-hand side of Eqn\,\ref{eqn:r_half}, while the cumulative flux inside $R_{1/2}$ is dominated by the compact-disc emissivity. In this case Eqn\,\ref{eqn:r_half} reduces to:
\begin{equation}
    \label{eqn:r_half_optUV}
    2\pi \int\limits_{R_{\rm in}}^{R_{1/2}} \epsilon_{\nu,{\rm disc}}(R)\,R\,\mathrm{d}R
    = \frac{1}{2}\left(L_{\nu,{\rm disc}} + L_{\nu,{\rm BLR}}\right).
\end{equation}
Increasing $L_{\nu,{\rm BLR}}$ relative to $L_{\nu,{\rm disc}}$ therefore increases the radius required to enclose half of the total observed flux. However, the magnitude of this increase is controlled not only by the BLR flux fraction, but also by the radial shape of the compact-disc emissivity profile, $\epsilon_{\nu,{\rm disc}}(R)$. A centrally concentrated emissivity profile will require a smaller increase in $R_{1/2}$ than a shallower profile for the same BLR flux fraction.
A key implication of Eq.\,\ref{eqn:r_half_optUV} is that, when $f_{\rm BLR}<0.5$ and the half-light radius remains inside the compact-disc region, the inferred size overestimation is set primarily by the combination of $f_{\rm BLR}$ and $\epsilon_{\nu,{\rm disc}}(R)$, rather than by the absolute BLR radius itself, provided that the BLR lies outside the compact continuum-emitting region. This explains why moderate BLR continuum fractions can produce noticeable, but not BLR-scale, increases in the inferred microlensing size. If the effect were simply a flux-weighted average between the compact disc and the BLR radius, even a BLR contribution of order $f_{\rm BLR}\sim0.3$ would imply an apparent size increase that scales directly with the much larger BLR radius. Instead, in the optical/UV regime relevant here, the inferred size remains governed by the cumulative compact-disc emissivity needed to enclose half of the total disc+BLR flux.

\rev{\subsection{Caveats}}

\rev{
There are a number of simplifying assumptions and caveats that should be kept in mind when interpreting these results.
}\\

\begin{enumerate}
    \item \rev{The BLR geometry and emission. We have assumed a highly simplified bi-conical BLR geometry with a single, constant gas density. In reality, the BLR geometry is likely to be more complex, containing a range of gas densities and thus ionisation states. The present work should therefore be considered as a first-order demonstration of how BLR emission can affect disc-size estimates inferred from microlensing.}

    \item \rev{Projection effects. We have assumed a close-to face-on source, as we are mainly interested in studying the central engine. As the inclination increases, our assumed circular (in the plane of the disc) accretion flow would be projected into an ellipse, which can reduce the effective half-light radius. Additionally, at high inclinations one expects material from the BLR, disc winds, and other large-scale-height structures to enter the line-of-sight, obscuring the central engine. This is not included in our model. For the simplified BLR geometry and covering fraction considered in this paper, the inclination at which one expects to intercept the BLR occurs at $\cos(i) \leq f_c=0.3$. However, this is likely an underestimate. The covering fraction in emission is predominantly set by the high-emissivity regions, which to first order trace gas density. If we interpret the BLR as originating from a disc wind \citep[e.g][]{Czerny11}, then the high-density, and therefore high-emissivity, regions will predominantly occur towards the base. There may still be regions at larger scale-heights that are not strongly emitting, yet would still impact the observations through absorption. This is evidenced in local Seyferts (e.g. NGC\,4151, NGC\,5548, Mrk\,817), which all show complex absorption signatures despite having moderate inclinations ($\cos(i) \sim 0.6$) \citep[e.g][]{Kara21, Zoghbi19, Dehghanian19b}. Hence, we do not recommend the use of our model for observer inclination of $\cos(i) \lesssim 0.7$.}

    \item \rev{Effective transverse velocity. We have assumed a single fiducial value of $v_{\rm eff}=600\,\mathrm{km\,s^{-1}}$ for all light-curve simulations, motivated by the peculiar-velocity scale inferred by \citet{Mediavilla2016}. However, the peculiar velocity represents only one contribution to the effective transverse velocity, which depends on the specific lensing system and can vary substantially from case to case \citep{Kochanek2004,Mosquera2011,Neira2020}. For a fixed monitoring duration, a lower effective transverse velocity corresponds to a shorter trajectory across the magnification map and therefore to less sampling of the caustic network. This reduces the amount of microlensing variability captured by the light curve and is expected to weaken the resulting source-size constraints.}

    \item \rev{Source size relative to the Einstein radius. We have assumed a fiducial black hole mass of $M=10^8 M_{\odot}$, as this corresponds to the typical black hole mass across the wider population seen with SDSS \citep{Wu22} For fixed dimensionless accretion parameters, increasing the black-hole mass also increases the physical scale of both the accretion flow and the BLR. The microlensing sensitivity therefore depends on the ratio between the characteristic source size and the Einstein radius, $r_{\rm source}/r_{\rm E}$. If the emitting region becomes much larger than $r_{\rm E}$, microlensing variability is strongly suppressed and the source size can no longer be accurately constrained, yielding primarily a lower limit.}
    
\end{enumerate}

\section{Summary and Conclusions}\label{conclusions}
In this work, we investigated how physically motivated AGN emissivity profiles affect quasar microlensing signatures and the inferred sizes of optical/UV continuum-emitting regions. We adapted the radially stratified {\sc agnsed} models based on \citet{Kubota18} to construct compact-disc emissivity profiles, and included reprocessed diffuse BLR continuum emission, as expected from \citet{Korista01,Korista19,Netzer22}, computed from the same ionizing SED. This allowed us to test how additional continuum-emitting components on larger spatial scales can introduce systematic biases in microlensing-based disc-size measurements.

We first demonstrated the impact of the different emissivity prescriptions by convolving each source model with both positive- and negative-parity source-plane magnification maps. The resulting distributions indicate the expected microlensing amplitudes. In general, models including diffuse BLR continuum emission predict narrower magnification distributions, corresponding to lower observed microlensing amplitudes and reduced variability. This behaviour arises because the larger-scale BLR emission smooths the small-scale caustic structure in the source-plane magnification maps. The strength of this effect is naturally wavelength dependent, since both the intrinsic compact-disc emissivity profile and the fractional BLR contribution to the SED vary with $\lambda$.

To assess the impact on observed microlensing studies more directly, we used our emissivity models to simulate {\sc cosmograil}-like light curves. This was done by drawing tracks across the unconvolved source-plane magnification maps and calculating the magnified luminosity at each point for each source model. The resulting ensemble of mock light curves was then used to recover source sizes from the microlensing variability statistics. 
The recovered size estimates tend toward the half-light radius of the source model used to generate the mock light curves, largely independently of the emissivity profile adopted in the forward modelling. This is consistent with the expectation that microlensing is primarily sensitive to the characteristic source size rather than to the detailed surface-brightness profile \citep{Mortonson2005}.

This result also demonstrates that larger-scale continuum-emitting material can bias compact-only microlensing size estimates. In particular, when light curves generated from the composite disc+BLR model are interpreted using compact-only source prescriptions, the recovered half-light radius corresponds to the effective size of the composite emission rather than to the true compact-disc size. This is consistent with the effect suggested by \citet{Fian2023_diffuse}, but here we demonstrate the recovered sizes directly using physically motivated AGN emissivity models. The resulting bias is conceptually similar to that invoked in continuum reverberation campaigns, where diffuse BLR emission can increase the observed continuum lag \citep{Korista19,Lawther18,Chelouche2019,Netzer22,Hagen24a}. In the microlensing case, however, the apparent size overestimation is not simply a flux-weighted average: it depends on both the fractional BLR contribution and the compact-disc emissivity profile at a given wavelength.

\rev{These simulations demonstrate the existence of a mechanism by which diffuse BLR continuum emission can bias compact-disc microlensing size estimates, but do not imply that this effect accounts for the full magnitude of the observed microlensing disc-size tension. Its importance in individual systems will depend on the wavelength-dependent diffuse-continuum contribution and on the spatial structure of the BLR.} 

Overall, our results show that neglecting additional continuum-emitting components can lead to biased microlensing size estimates. In particular, diffuse BLR continuum emission increases the effective half-light radius of the observed optical/UV continuum and dilutes the microlensing signal from the compact accretion disc. This suggests that part of the apparent excess in microlensing-inferred quasar disc sizes may arise from interpreting composite disc+BLR emission as emission from a single compact accretion-disc component.

\section*{Data availability}
No new observational data are presented in this paper. The model code, {\sc AGNmap}, is available via GitHub at \url{https://github.com/scotthgn/AGNmap}.\\

\begin{acknowledgements}
      \rev{We thank the anonymous referee for their constructive comments, which imprved the clarity and robustness of the paper.} S.H. acknowledges support from the IFPU fellowship scheme. C.F. has received funding from the European Union's Horizon Europe research and innovation programme under the Marie Sk{\l}odowska-Curie COFUND Postdoctoral Programme, grant agreement No. 101081355 (SMASH), and from the Republic of Slovenia and the European Union through the European Regional Development Fund. Views and opinions expressed are, however, those of the author(s) only and do not necessarily reflect those of the European Union or the European Research Executive Agency. Neither the European Union nor the granting authority can be held responsible for them. \\
      Author Contributions: S.H. constructed the physically motivated accretion-disc models and performed the {\sc cloudy} calculations used to model the diffuse BLR continuum emission. S.H. also converted the disc and BLR emission components into two-dimensional source emissivity maps, performed the model convolutions with the magnification maps, and generated the simulated mock light curves. C.F. generated the source-plane microlensing magnification maps, characterized the statistical properties of representative microlensing light curves, and performed the source-size inference analysis using the mock data. Both authors contributed to the interpretation of the results and to writing the manuscript.
\end{acknowledgements}

\bibliographystyle{aa}
\bibliography{Refs}

\begin{appendix}

\onecolumn
\section{Biconical BLR geometry}\label{app:blr_goem}
\label{app_sec:BLR_geom}
\begin{figure*}[h]
    \centering
    \includegraphics[width=0.7\textwidth]{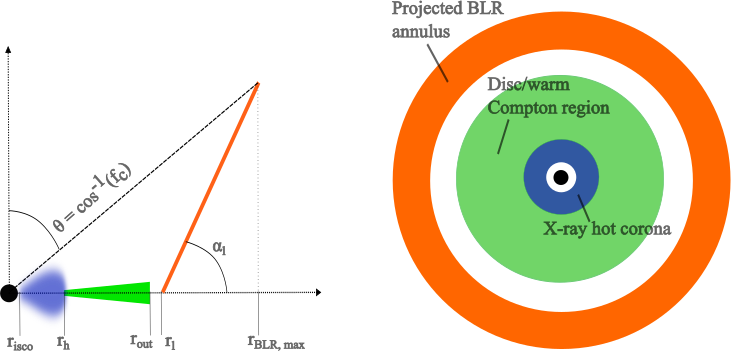}
    \caption{
    Sketch of the BLR geometry adopted in this work. \textbf{\textit{Left:}} the BLR is modelled as a single bicone launched at radius $r_l$ and opening at an angle $\alpha_l$ with respect to the disc plane, subtending a total covering fraction $f_c$. These parameters define the outer radius $r_{\rm BLR,max}$ measured in the disc plane. This geometry is used to compute the total diffuse BLR emission and follows the prescriptions of \citet{Hagen23a} and \citet{Hagen24a}. \textbf{\textit{Right:}} for the microlensing calculation, the BLR emission is projected onto the disc plane, producing an annular emitting region between $r_l$ and $r_{\rm BLR,max}$.
    }
    \label{app_fig:geom_sketch}
\end{figure*}

Following \citet{Hagen23a}, we model the BLR as a single bi-conical geometry, parametrized by the launch radius $r_l$, the launch angle $\alpha_l$, and covering fraction $f_c = \Omega/4\pi$ (see the left panel of Fig.\,\ref{app_fig:geom_sketch}). These parameters define the outer extent of the BLR. In cylindrical coordinates, this gives
\begin{equation}
    r_{\rm{BLR, max}} = \frac{r_l \tan(\alpha_l)}{\tan(\alpha_l) - \tan(\frac{\pi}{2} - \theta)}
\end{equation}
where $\theta = \cos^{-1}(f_c)$. We use {\sc cloudy} v.25.00 \citep{Ferland17, Gunasekera25} to estimate the BLR emission contribution to the SED. For simplicity, we assume a BLR with uniform gas density and compute a single {\sc cloudy} model to estimate the total BLR emission at each energy, $L_{\nu, \rm{BLR}}$ (see Section\,\ref{sec:blr_geom}). Since the observer is assumed to view the system along the bi-cone axis, the emission from the far side of the BLR is obscured by the disc. We therefore take $L_{\nu,\rm BLR}$ to be one half of the total emission predicted by {\sc cloudy}.

To estimate the microlensing signature, we project the BLR emission onto the disc plane. For simplicity, we assume that the BLR emission is isotropic and uniformly distributed over the BLR surface. This approximation leads to a simple projected top-hat emissivity profile, confined to the annulus between the launch radius $r_l$ and the maximum BLR radius $r_{\rm BLR,max}$:
\begin{equation}
    \epsilon_{\nu, \rm{BLR}} = \frac{L_{\nu, \rm{BLR}}}{\pi R_G^2 (r_{\rm{BLR, max}}^2 - r_l^2)}
\end{equation}

\section{Size-recovery procedure overview}\label{size_recovery}
\begin{figure*}[h]
    \centering
    \includegraphics[width=\textwidth]{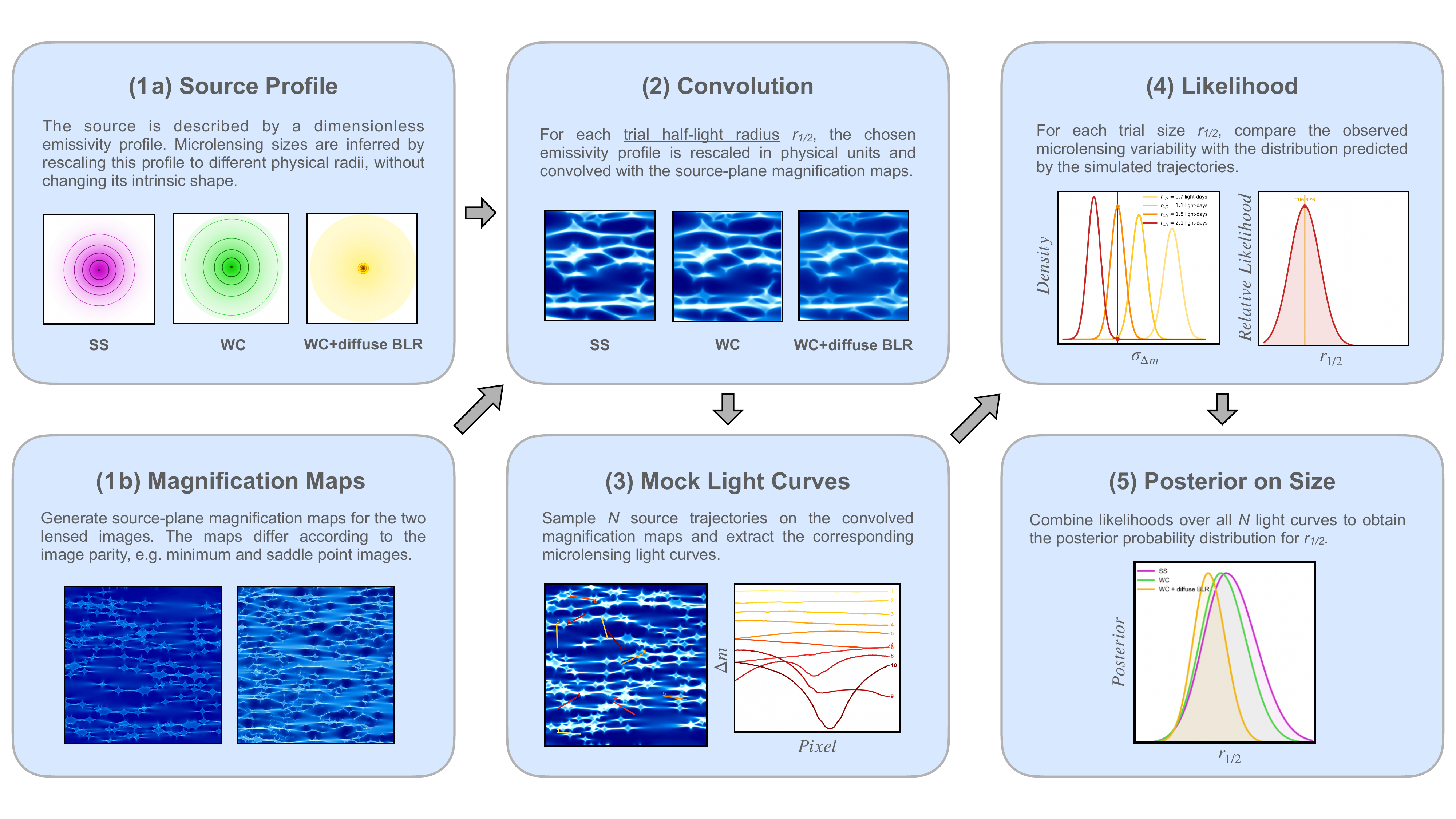}
    \caption{Schematic overview of the microlensing size-inference procedure. First (1\,a), a dimensionless source emissivity profile is adopted for each physical model, here illustrated for a Shakura–Sunyaev disc (SS), a warm-Comptonisation disc (WC), and a warm-Comptonisation disc including diffuse BLR continuum emission (WC+diffuse BLR). Source-plane microlensing magnification maps are generated for the two lensed images (1\,b), accounting for their different image parities. For each trial half-light radius $r_{1/2}$, the chosen source profile is rescaled in physical units and convolved with the magnification maps (2). Mock microlensing light curves are then extracted by sampling $N$ source trajectories across the convolved maps (3). The observed microlensing variability is compared with the distribution predicted by the simulated trajectories to evaluate the likelihood of each trial size (4). Combining the likelihoods over all light curves yields the posterior probability distribution for $r_{1/2}$ (5).
}
    \label{size_recovery_schematic}
\end{figure*}

In Fig.~\ref{size_recovery_schematic}, we show a schematic overview of the size-recovery procedure described in Section~\ref{sec:size_recovery_statistics}. The figure illustrates how trial source sizes are generated from a fixed emissivity profile, convolved with the magnification maps, sampled along mock trajectories, and compared with the variability measured from the mock microlensing light curves to infer the most likely half-light radius.

\section{Recovered disc sizes}\label{recovered_sizes}
As a consistency check, we repeated the size inference on mock light curves generated and fitted with the same source model, namely SS$\rightarrow$SS, WC$\rightarrow$WC, and WC+BLR$\rightarrow$WC+BLR. The recovered median half-light radii are generally close to the input values (see Table~\ref{tab:self_recovery}), although they are systematically slightly larger, with an average recovered-to-true ratio of $\sim 1.16$ across all bands, parities, and source models. Since this is a simplified size-recovery test, it should be regarded as a baseline validation rather than a full systematic-error analysis. Even so, the resulting offset is too small to explain the much larger size excesses obtained when BLR-contaminated light curves are interpreted with compact-only source models.
\begin{table}
\caption{Recovered sizes from same-model mock tests.}
\label{tab:self_recovery}
\centering
\setlength{\tabcolsep}{15pt}
\renewcommand{\arraystretch}{1.4}
\begin{tabular}{lcccc}
\hline\hline
Emissivity Profile & $\lambda$ 
& \multicolumn{3}{c}{$r_{1/2}$ (light-days)} \\
\cline{3-5}
& ($\mathrm{\AA}$) 
& Input 
& Positive parity 
& Negative parity \\ \hline
Shakura-Sunyaev 
& 1650 & 0.29 & $0.33_{-0.03}^{+0.03}$ & $0.35_{-0.04}^{+0.03}$ \\
& 3250 & 0.61 & $0.69_{-0.06}^{+0.06}$ & $0.69_{-0.07}^{+0.07}$ \\
& 7500 & 1.45 & $1.73_{-0.13}^{+0.13}$ & $1.56_{-0.12}^{+0.12}$ \\
\hline
Warm Comptonisation
& 1650 & 0.39 & $0.45_{-0.04}^{+0.04}$ & $0.46_{-0.04}^{+0.05}$ \\
& 3250 & 0.74 & $0.89_{-0.08}^{+0.07}$ & $0.84_{-0.07}^{+0.08}$ \\
& 7500 & 1.48 & $1.78_{-0.13}^{+0.14}$ & $1.61_{-0.12}^{+0.12}$ \\
\hline
Warm Comptonisation+BLR 
& 1650 & 0.40 & $0.47_{-0.06}^{+0.08}$ & $0.48_{-0.06}^{+0.08}$ \\
& 3250 & 1.12 & $1.40_{-0.10}^{+0.10}$ & $1.26_{-0.09}^{+0.10}$ \\
& 7500 & 2.35 & $2.75_{-0.23}^{+0.17}$ & $2.56_{-0.25}^{+0.30}$ \\
\hline
\end{tabular}
\end{table}

\rev{\section{Sensitivity to the kernel-density estimate}\label{app:kde_sensitivity}}

\begin{figure*}
    \centering
    \includegraphics[width=0.9\textwidth]
    {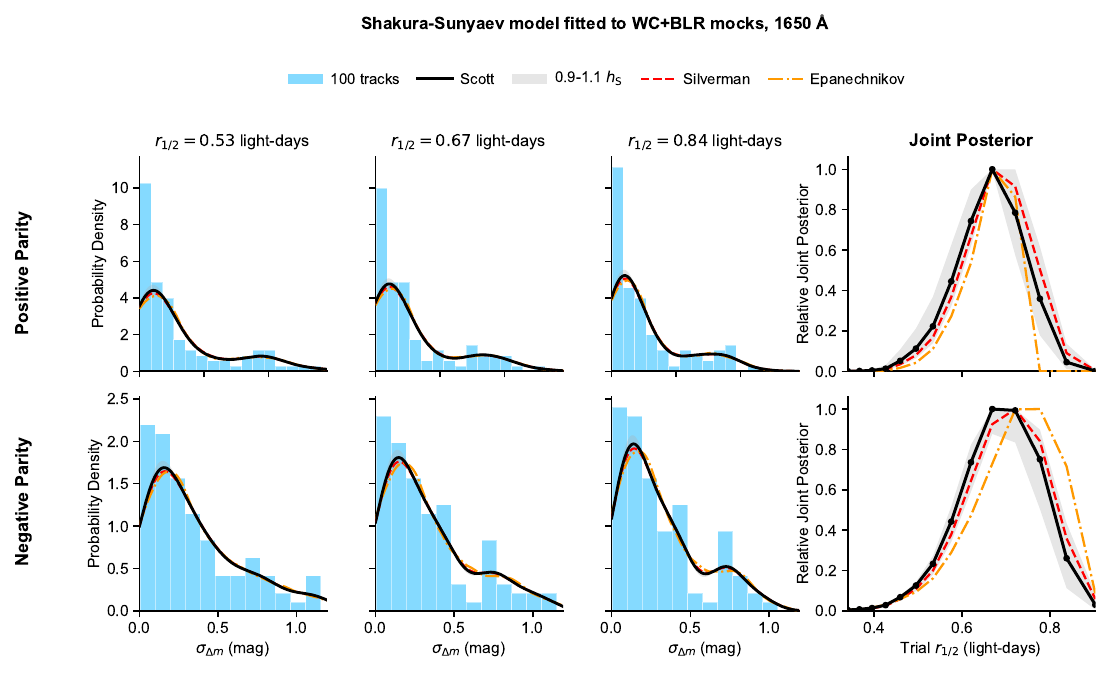}
    \caption{\rev{Sensitivity of the variability distributions and recovered half-light radii to the KDE prescription. The first three columns show the distributions of $\sigma_{\Delta m}$ obtained from 100 simulated trajectories for a Shakura-Sunyaev emissivity profile in the 1650\,\AA\ band at three representative trial half-light radii. The upper and lower rows correspond to positive- and negative-parity magnification maps, respectively. Histograms show the simulated samples, black curves show the fiducial Gaussian KDE using Scott's bandwidth, and grey regions show the range obtained by varying the bandwidth between $0.9h_{\rm S}$ and $1.1h_{\rm S}$. Red dashed and
    orange dash-dotted curves show the Gaussian KDE using Silverman's bandwidth and the variance-matched Epanechnikov KDE, respectively. The final column shows the corresponding joint half-light-radius posteriors obtained by fitting the warm-Comptonisation+BLR mock light curves with the Shakura-Sunyaev emissivity profile. Each posterior is normalized to its maximum value.}}
    \label{fig:kde_sensitivity}
\end{figure*}

\rev{For each trial half-light radius and assumed emissivity model, the distribution of the light-curve variability statistic, \mbox{$p(\sigma_{\Delta m}\mid r_{1/2},M)$}, is estimated from the $N_{\rm LC}=100$ simulated trajectories using a one-dimensional kernel-density estimate (KDE),}
\rev{\begin{equation}
    \widehat{p}_{\rm KDE}(\sigma_{\Delta m})
    =
    \frac{1}{N_{\rm LC}h}
    \sum_{j=1}^{N_{\rm LC}}
    K\!\left(
    \frac{\sigma_{\Delta m}-\sigma_{\Delta m,j}}{h}
    \right),
\end{equation}}
\rev{where $K$ is the kernel function and $h$ is the bandwidth. Our fiducial analysis adopts a Gaussian kernel with the bandwidth determined using Scott's rule \citep{Scott1992},
$h_{\rm S}=N_{\rm LC}^{-1/5}s$, where $s$ is the sample standard deviation of the simulated $\sigma_{\Delta m}$ values at a given trial half-light radius. Since the number of trajectories is finite and the KDE depends on both the adopted kernel and bandwidth, we test the sensitivity of the inferred source sizes to these choices. Figure~\ref{fig:kde_sensitivity} shows a representative example using a Shakura-Sunyaev emissivity profile for the size recovery in the 1650\,\AA\ band. The first three columns show the distributions of $\sigma_{\Delta m}$ at three trial half-light radii spanning the region relevant to the recovered posterior. Each histogram contains the same 100 simulated trajectories used in the size-inference analysis. The histogram binning is chosen using the Freedman-Diaconis rule \citep{FreedmanDiaconis1981}, with the bin width held fixed across the three trial radii for each image parity. The black curve shows the
fiducial Gaussian KDE obtained using Scott's bandwidth, while the grey region shows the range obtained by varying the bandwidth between $0.9h_{\rm S}$ and $1.1h_{\rm S}$. For comparison, we also show a Gaussian KDE using Silverman's bandwidth prescription \citep{Silverman1986} and a variance-matched Epanechnikov kernel \citep{Epanechnikov1969}. The final column shows how these choices propagate into the joint posterior on the half-light radius. For each parity, the KDE at every trial radius is evaluated at the variability statistics measured from the 100 warm-Comptonisation+BLR mock light curves. The corresponding posterior is}
\rev{\begin{equation}
    p\!\left(r_{1/2}\mid\{\sigma_i\},M\right)
    \propto
    p(r_{1/2})
    \prod_{i=1}^{N_{\rm LC}}
    \widehat{p}_{\rm KDE}
    \!\left(\sigma_i\mid r_{1/2},M\right),
\end{equation}}
\rev{where $M$ denotes the assumed Shakura-Sunyaev emissivity profile and $p(r_{1/2})\propto r_{1/2}^{-1}$ is the adopted log-uniform prior. For display purposes, each posterior is normalized to its maximum value.
For the positive-parity map, the fiducial Scott KDE gives a posterior median of $r_{1/2}=0.67^{+0.08}_{-0.08}$ light-days. Varying the bandwidth to $0.9h_{\rm S}$ and $1.1h_{\rm S}$ changes the median by $-4.1\%$ and $+3.8\%$, respectively. Using Silverman's bandwidth changes the median by $+2.3\%$, while the variance-matched Epanechnikov kernel changes it by $+0.5\%$. For the negative-parity map, the fiducial result is $r_{1/2}=0.69^{+0.09}_{-0.10}$ light-days. The corresponding shifts in the median are $-3.1\%$ and $+3.0\%$ for the $0.9h_{\rm S}$ and $1.1h_{\rm S}$ bandwidths, respectively, $+1.8\%$ for Silverman's prescription, and $+5.9\%$ for the Epanechnikov kernel. In all cases, the changes induced by the KDE prescription are smaller than the statistical width of the corresponding posterior and do not affect the qualitative size-recovery result. We therefore conclude that, for this representative case, the inferred half-light radius is not strongly sensitive to the particular standard KDE prescription adopted in the fiducial analysis.}

\end{appendix}
\end{document}